
\magnification = 1200
\overfullrule = 0 pt
\baselineskip = 16 pt
\hsize = 14.5 truecm
\vsize = 22. truecm

\font\cp = cmsl9 scaled \magstep0
\font\cpp = cmr9 scaled \magstep0

\def\diracij{\delta_{{\bf i}, {\bf j}}}

\def\DD{D_{\bf i}}
\def\AAi{A_{\bf i}}
\def\AdAi{A_{\bf i}^{\dagger}}
\def\AAj{A_{\bf j}}
\def\NNi{N_{\bf i}}

\def\CcC{{\hbox{\tenrm C\kern-.45em{\vrule height.67em width0.08em depth-.04em
\hskip.45em }}}}
\def\RrR{{\hbox{\tenrm I\kern-.17em{R}}}}
\def\HhH{{\hbox{\tenrm {I\kern-.18em{H}}\kern-.18em{I}}}}
\def\DdD{{\hbox{\tenrm {I\kern-.18em{D}}\kern-.36em {\vrule height.62em
width0.08em depth-.04em\hskip.36em}}}}
\def\ZzZ{{\hbox{\tenrm Z\kern-.31em{Z}}}}
\def\IiI{{\hbox{\tenrm I\kern-.19em{I}}}}
\def\NnN{{\hbox{\tenrm {I\kern-.18em{N}}\kern-.18em{I}}}}

\hrule height 0.0 pt
\nopagenumbers
\line{\hfill POLFIS-TH.03/92}
\vskip 2. truein

\centerline{\bf FERMI--LINEARIZATION SCHEME FOR ITINERANT}
\vskip .15 truein
\centerline{{\bf ELECTRONS WITH CLIFFORD VARIABLES}}
\vskip .4 truein
\centerline{\sl Arianna Montorsi, and Alessandro Pelizzola}
\vskip 0.5cm

\centerline{Dipartimento di Fisica and Unit\'a INFM,}
\centerline{Politecnico di Torino, I-10129 Torino, Italy}

\vskip .7 truein
\centerline{\bf Abstract}
\vskip .12 truein
{\cp {We propose here an alternative interpretation of the fermi-linarization
approach to interacting electron systems, based on the requirement that the
coefficients of the linearized operators are Clifford-like variables, whose
anticommutator equals an unknown constant $c$. We apply the approximation to
the Falicov-Kimball model, explicitly solving the self-consistency equation for
the unknown, which turns out to behave as an order parameter. We discuss its
relation with a metal-insulator transition and some thermodynamical quantities.
In particular we show that our approximation in the $T=0$ limit reproduces
exactly the Gutzwiller results for the Hubbard model. }}
\vskip .4 truein
\line{P.A.C.S. \# 05.30.Fk 71.10.+x \hfill}
\vfill\eject
\pageno 2
\footline{\hss \tenrm -- \folio $\,$ -- \hss}
\hrule height 0 pt
\vskip 2 truecm
{\bf 1. Introduction}
\medskip
Both systems of itinerant interacting electrons on an infinite lattice, and
simple systems of few electrons interacting with bosons, are generally
described by hamiltonians whose dynamical algebra is infinite dimensional.
Various approximation techniques have been developed in order to deal with such
systems. In particular, in a set of recent papers$^{[1-3]}$ an approximation
scheme was proposed, referred to as fermionic linearization scheme, which can
be applied to a generic many-fermion hamiltonian. It consists in replacing in
the hamiltonian certain bilinear products of (sums of) electron creation or
annihilation operators, say $a_1$ and $a_2$, by terms linear in some fermion
operator $f$ multiplied by appropriate Grassmann-like coefficients $\theta$,
$$
a_1 a_2 + a_2^\dagger a_1^\dagger \sim \theta f + f^\dagger \bar\theta \quad ,
\eqno{(1)}
$$
where $\{\theta,\bar\theta\}=0$, and $\{\theta,f\}= \{\bar\theta,f\}= 0=
\{\theta,a_i\}=\{\bar\theta,a_i^\dagger\}$ ($i=1,2$). The anticommutation
relations of $f$ and $f^\dagger$ are uniquely determined by $a_1$ and $a_2$,
and depend on the problem studied. The fact that both the operators
$f,\,f^\dagger$ and the Grassmann coefficients $\theta,\, \bar\theta$ satisfy
anticommutation relations guarantees that the bilinear products on the $r.h.s.$
of (1) have the same 'statistics' of the bilinear operators at the $l.h.s.$.

Once substitution (1) is performed, the scheme allows one to obtain the
spectrum of the linearized hamiltonian -- after recognizing that the
'effective' model has a dynamical algebra which is a ${\ZzZ}_{2}$-graded
algebra --  via an inner automorphism of the algebra itself (which generalizes
the customary Bogolubov rotation).

In the present note, we propose a new view of the fermionic linearization
scheme, which consists in requiring that the variables $\theta,\bar\theta$
satisfy a Clifford-like instead of the Grassman-like algebra. More precisely,
we set
$$
\{\theta,\,\bar\theta\} = c^2 \quad, \quad c\in \RrR \quad , \eqno{(2)}
$$
with $c$ an undeterminate, to be defined for each specific problem. Notice that
the requirement (2) on the $\theta$'s implies that the dynamical algebra of the
linearized model is no longer graded, but simply a Lie algebra. In other words,
we require that the $\theta$'s behave as operators rather than anticommuting
numbers. Indeed, by inspection of (1) one can easily verify that, in the simple
case in which $a_1$ and $a_2$ are single electron operators, and
$\{f,f^\dagger\}=1$, eqn. (1), with $c=1$, maps a two-electrons operator into
another two-electrons operator, hence the approximation of the {\it r.h.s.}
term of (1) becomes exact. In general, this is not true, and a value of $c$ has
to be determined self-consistently according to eq. (1). The self-consistency
equations reconduct then the exact results for the linearized model to
approximate (mean-field like) results for the original hamiltonian.

In the following, we will use this approximation to the solution of the
Falicov-Kimball model. The latter gives a very simplified description of
a system of itinerant fermions interacting only locally. In this case,
prescription (1) is applied to the itinerant part of the hamiltonian,
reducing it to an effective single-site operator, while it leaves
unchanged the interaction term. The resulting approximation thus in principle
goes well beyond the standard weak-coupling mean-field theory, and
indeed it turns out to be capable of describing a metal-insulator transition.

Let us observe that the approximation (2) was already
used in a different context$^{[4]}$ with a fixed value for $c$,
{\sl i.e.} $c=1$.
\bigskip
{\bf 2. The Falicov-Kimball model}
\medskip
The Falicov-Kimball model$^{[5]}$ provides a very simple
description of large systems of itinerant interacting fermions,
by considering two different species of electrons (say with up and down spin)
on a lattice $\Lambda$, one of each itinerates on $\Lambda$, the electrons
with opposite spins being fixed at their sites, and assuming that the electrons
interact only via an on-site Coulomb repulsion term.
The grand-canonical hamiltonian reads
$$
H_{FK} = - \mu_n \sum_{\bf i} \NNi -\mu_d \sum_{\bf i}\DD - 2 t \sum_{<{\bf i},
{\bf j}>} \AdAi \AAj + U \sum_{\bf i} \NNi\DD \quad , \eqno{(3)}
$$
where $t>0$ is the hopping amplitude, and $U>0$ is the local electron-electron
repulsion. $\AdAi , \AAi$ are operators which create and annihilate the
itinerant electrons ($ \{ \AAi , \AAj \} = 0 \,$ , $\, \{ \AdAi , \AAj \} =
\diracij {\IiI}$, $\, \NNi\doteq \AdAi \AAi$, ${\bf i} , \, {\bf j} \in
\Lambda$), and $< {\bf i} , \, {\bf j} >$ stands for non-oriented nearest
neighbours ({\sl n.n.}) in $\Lambda$. Moreover $\DD$ is the number operator of
the non-itinerant electrons. As the operators  $\sum_{\bf i} \NNi$ and
$\sum_{\bf i} \DD$ both commute with the hamiltonian, the chemical potentials
$\mu_n$ and $\mu_d$ allow to fix the average number of electrons of the two
species.

The Falicov-Kimball model was introduced for studying the metal-insulator
transition in transition metal oxides, and can be considered as a simplified
version of the Hubbard model$^{[6]}$. The exact statistical mechanical solution
for the model described by $H_{FK}$ is known only for large dimensions$^{[7]}$.
However, a few general theorems are known$^{[8]}$ for the symmetric (or
neutral)
case $\mu_n=\mu_d=\displaystyle {U\over 2}$, and in particular an Ising-like
phase transition is expected for dimension $D\geq 2$ at some critical
temperature, whose value should vanish both for small and large $U$. Moreover,
there are a number of investigations of the ground state phase diagram in
dependence on the configuration of fixed spins$^{[9]}$. Also, a strong-coupling
($U>>t$) thermodynamic mean-field theory -- based on the $D=\infty$ exact
solution -- was proposed$^{[10]}$.

The fermionic linearization approach$^{[1-3]}$, mentioned in the introduction,
provides as well a powerful approximation scheme for Hubbard-like models in the
strong coupling limit. In fact it treats in an exact way the Coulomb
interaction term, whereas it acts only on the hopping term. Let us write the
latter as
$$
\sum_{<{\bf i},{\bf j}>} \AdAi \AAj  =
{q\over 2}\sum_{{\bf i}} \left (\Theta_{\bf i}^\dagger \AAi +
\AdAi \Theta_{\bf i}\right ) \quad, \eqno{(4)}
$$
with $\displaystyle{\Theta_{\bf i}\doteq {1\over{q}}\sum_{{\bf j} n.n.
{\bf i}} \AAj}$, $q$ denoting the number of nearest neighbours of a site
in $\Lambda$.
Of course, the operators $\Theta_{\bf i}$ have non trivial anticommutation
relations among themselves as well as with the $\AAi$'s. On the other
hand, in [1-3] the $\Theta_{\bf i}$'s were approximated by variables
$\theta_{\bf i}$'s anticommuting among themselves as well as with the fermion
operators, {\sl i.e.} $\bigl \{\theta_{{\bf i}},\theta_{{\bf j}} \bigr\} = 0 =
 \bigl \{\bar \theta_{{\bf i}},\theta_{{\bf j}} \bigr\} $,
$\bigl\{\bar\theta_{{\bf i}},\AAj \bigr\} = 0 =
\bigl\{\theta_{{\bf i}},\AAj \bigr\}$,$ \forall {\bf i}, {\bf j} \in
\Lambda$. The former prescription is exact only for ${\bf i}$ and ${\bf j}$
far enough, depending on the lattice $\Lambda$, whereas it is definitely
too simple for ${\bf i}$ and ${\bf j}$ coinciding or having nearest neighbours
in common.

Here we propose to improve the fermionic approximation scheme by replacing the
operators $\Theta_{\bf i}$ by variables $\theta_{\bf i}$ still anticommuting
with the fermion operators, and satisfying the
following algebra (which is a straightforward generalization of (2))
$$
\{\theta_{\bf i},\bar\theta_{\bf j}\} = c^2 \,\delta_{{\bf i}, {\bf j}} \quad ,
\quad \{\theta_{\bf i},\theta_{\bf j}\} = 0 \quad . \eqno{(5)}
$$
Once the above approximation is inserted in (3), one obtains a reduced
hamiltonian ${\cal H}_{FK}$ which is a sum over lattice
sites of single-particle hamiltonians, ${\cal H}_{\bf i}$, commuting
with each other,
$$
{\cal H}_{\bf i} = -\mu_n \NNi - \mu_d \DD - t q {c} (\bar \eta_{\bf i}
\AAi +\AdAi \eta_{\bf i}) + U \NNi \DD, \quad, \eqno{(6)}
$$
with $\displaystyle{\eta_{\bf i}\doteq {\theta_{\bf i}\over{c}}}$ ,so that
$\{\bar\eta_{\bf i}, \eta_{\bf j} \} =\delta_{{\bf i},{\bf j}}$.

The $\DD$'s are to be considered as classical, Ising-like, variables, whose
two possible eigenvalues 0 and 1 label two orthogonal projections of
${\cal H}_{\bf i}\doteq {\cal H}_{\bf i}^{(0)}\oplus {\cal H}_{\bf i}^{(1)}$.
The problem of finding the spectrum of hamiltonian (3) is thus reduced,
after linearization, to that of diagonalizing the local effective hamiltonian
${\cal H}_{\bf i}^{(\DD)}$. In order to do it, one should
first identify the dynamical algebra, ${\cal A}$, of (6);
it is easily verified that the latter coincides with $u(2)$, generated by
$$
{\cal A}\equiv u(2)=\left\{\NNi\pm \bar\eta_{\bf i}\eta_{\bf i}\, ; \,
\bar \eta_{\bf i} \AAi \pm \AdAi \eta_{\bf i}\right\} \quad . \eqno{(7)}
$$
The transformation which rotates the hamiltonian into its diagonal form
$\tilde{\cal H}_{FK}$ is then obtained by acting on ${\cal H}_{\bf i}^{(\DD)}$
with
$\exp{({\rm ad} Z)}\doteq \displaystyle{\sum_{n=0}^{\infty} {1\over{n!}}}
[Z,[Z, \dots,[Z, \bullet]\dots ]]$,
where $Z$ is an appropriate skew-hermitian non-Cartan
element of ${\cal A}$, $Z= p (\bar \eta_{\bf i} \AAi - \AdAi \eta_{\bf i})$.
It is easily verified that the choice $p=\displaystyle{
{\rm arctg}{{2 \tau}\over{U\DD-\mu_n}}}$, with $\tau={c} q t$, implies
$$
\tilde{\cal H}_{FK} = {1\over 2}\left\{\epsilon_{\bf i} (\bar\eta_{\bf i}
\eta_{\bf i} + \NNi) \pm \sqrt{\epsilon_{\bf i}^2 + 4 \tau^2} (\bar\eta_{\bf i}
\eta_{\bf i} - \NNi) \right \} - \mu_d \DD
\, , \eqno{(8)}
$$
with $\epsilon_{\bf i}\doteq U\DD-\mu_n$; $\tilde{\cal H}_{FK}$ is manifestly
diagonal.

The result (8) is also interesting from the point of view of statistical
mechanics, in that the partition function ${\cal Z}$ is immediately obtained
from (8) as
$$
{\cal Z}=\sum_{\NNi,\DD, \bar\eta_{\bf i}\eta_{\bf i} =0,1} \exp {\left
(-\beta \tilde{\cal H}_{FK} \right)} \quad . \eqno{(9)}
$$
Predictions for physical quantities can then be
obtained from ${\cal Z}$ once the average numbers of electrons of the two
species are fixed through the chemical potentials, according to
$$
\eqalign{n&\doteq <\NNi> = {1\over\beta {\cal Z}}{\partial {\cal Z}\over
\partial \mu_n} \quad ;\qquad (10.1)\cr
d&\doteq <\DD> = {1\over\beta {\cal Z}}{\partial {\cal
Z}\over \partial \mu_d} \quad , \qquad (10.2) \cr}
$$
where $<\bullet>$ stays for the thermodynamical average in the Gibbs ensemble
of operator $\bullet$ ({\sl i.e.} $<\bullet> = {\cal Z}^{-1}
\sum_{\NNi,\DD ,\bar\eta \eta =0,1}\bullet \exp
{(-\beta{\cal H}_{\bf i}^{(\DD)})}$ $\equiv {\cal Z}^{-1}$
$\sum_{\NNi,\DD ,\bar\eta \eta =0,1}\exp{\rm (ad Z)}(\bullet) \exp{(- \beta
\tilde{\cal H}_{FK}})$  ).

Moreover, in order to have quantitative predictions, a numerical value for
$c$ has still to be self-consistently determined. Indeed, the prescription of
substituting in the hopping term the
$\Theta_{\bf i}$ operators with the $\theta_{\bf i}$'s can be implemented once
more in (4), giving rise to the following self-consistency equation,
$$
<\bar\eta \AAi + \AdAi \eta> = 2{c} <\bar\eta \eta> \quad , \eqno{(11)}
$$
in which we have assumed translational invariance of the lattice, implying
$\eta_{\bf i} \equiv \eta, \forall {\bf i} \in \Lambda$.

The three equations (10.1), (10.2), and (11) have interesting features.
First of all, we notice that (10.2)
can be solved explicitly for $\mu_d$, and gives the result
$$
\exp {\beta\mu_d} = {d\over{1-d}} {\displaystyle{1+e^{\beta\mu_n}
+2e^{\beta\mu_n\over 2}\cosh{\beta\over 2}\sqrt{\mu_n^2 +4\tau^2}}\over
\displaystyle{1+e^{\beta(\mu_n-U)}+2
e^{{\beta(\mu_n-U)\over 2}}\cosh{\beta\over2}\sqrt{(\mu_n-U)^2
+4\tau^2 }}} \quad . \eqno{(12)}
$$
Moreover, it is easy to check that eq. (11) always factorizes a solution $c=0$,
which correspond to the insulating behavior. Besides this solution, in general
the system formed by (10.1)-(11), with $\mu_d$ given by (12), is highly
non-linear, and must be dealt with numerically. It turns out that it has
different non-zero solutions. The physical one is to be chosen as that which
minimizes the Gibbs free-energy $f$, $\displaystyle{f=-{1\over\beta} \ln Z}$.
In the next section we shall discuss the results of the numerical analysis, as
well as the analytical results which can be obtained in some limiting cases.
\bigskip
{\bf 3. Results and discussion}
\medskip
In figure 1. we report the mean-field parameter ${c}$ vs. temperature
$\displaystyle{{{k T}\over {q t}}}$, at half-filling and for the symmetric
case $n=d={1\over 2}$. In
this case it is easy to check that the solution to (10.1)--(10.2) is
$\displaystyle {\mu_n=\mu_d={U\over 2}}$. ${c}$ is plotted for different $U$
values, and exhibits a typical order-parameter like behavior. For $U=0$
(non-interacting case) it rises from zero, in the high-temperature regime, to
one, at $T=0$. For generic $U\leq 4 q t$, it is possible to show rigorously
that, in the limit $T\rightarrow 0$, ${c}$ reaches a value ${c_0}$ given by
$$
c_0^2 = 1-{1\over{16}}\tilde U^2 \quad , \eqno{(13)}
$$
where $\displaystyle{\tilde U\doteq {U\over{q t}}}$. This suggests that the
value $c=1$ used in [4] is correct at half filling, only in a
low-temperature non-interacting regime or for $D=\infty$. On the contrary,
for $U>4 q t$, the only solution to (11) is $c=0$.

The expression (13) for ${c_0}$ clarifies the physical meaning of
the parameter ${c}$. Indeed, recalling that on a hypercubic lattice $q$ is
twice the dimension of the lattice, eq. (13) reproduces exactly the
Gutzwiller result$^{[11]}$ for the discontinuity in the single particle
occupation number at the Fermi surface, obtained for the conventional Hubbard
model when $T=0$. This is not surprising as, on the one hand, the Gutzwiller
result for the Hubbard model was obtained in fact by neglecting the kinetic
energy of one species of electron, thus in an approximation very similar to
that at the basis of the Falicov-Kimball model. On the other hand, according to
eqs. (4), (5), and (11), at half filling $c$ coincides with the expectation
value of the hopping term, and hence is related to the discontinuity in its
Fourier transform.

Notice that when $U=0$ then $c_0=1$, and the ground state has all the electrons
below the Fermi level. For any $c\neq 0$, the ground state has some electrons
above the Fermi level, but the gap is still there, and, according to eq. (5),
the generic lattice site on which one has confined the linearized hamiltonian
is
still interchanging fermions with the rest of the lattice. When $c_0=0$ on the
other hand, the gap in the density of states disappears, and at half-filling we
have exactly one electron per site. In this case,  the remaining of the lattice
behaves as a system of scorrelated 'average' fermions ({\sl i.e.} as
if they were frozen at their own sites) and we are in presence of an
insulating phase.

The above analysis suggests that $c$ could be able to describe the transition
from a conducting to an insulating
state. Indeed, again in agreement with the Gutzwiller result, at $T=0$ we find
that the double occupancy expectation value, ${\cal P}\doteq <\NNi \DD>$,
vanishes precisely at $\tilde U = 4$. Explicitly, analytic calculation shows
that
$$
{\cal P} =\Biggl \{ \matrix{ \displaystyle{{1\over 4}\left (1 - {\tilde U\over
4}\right)}&\quad {\rm for} \,\tilde U\leq 4\cr
0 &\quad {\rm otherwise}} \quad . \eqno{(14)}
$$
It is worth noticing that the result (14) coincides with the exact result both
in the limit $\tilde U=0$ and in the limit $\tilde U>>1$.

A deeper analysis of figure 1. shows that  the transition from
non-zero to vanishing $c$ is of different order depending on the value of
$\tilde U$. Indeed, by requiring that (11) vanishes also around $|c|=0$, one
can verify
that there exists a tri-critical point at $\tilde U = U_t$, where $U_t$ is
solution of
$$
\tanh {{U_t}\over\displaystyle{2 \left (1-{{U_t^2}\over 8}\right )}}=
{U_t\over 2} \quad . \eqno{(15)}
$$
One finds a numerical value $U_t \simeq 1.845$. For $\tilde U$ smaller than
$U_t$ the transition is second order, and
the critical temperature is found analytically as the solution $T_c$ of
the following equation (obtained by requiring that (11), upon factorizing
the $c=0$ solution, still vanishes for $c=0$):
$$
\tanh {\tilde U\over{4 \Theta_c}} = \tilde {U\over 2} \quad, \eqno{(16)}
$$
with $\displaystyle{\Theta_c\doteq {{k T_c}\over{q t}}}$, and $k$ the
Boltzmann constant.
On the other hand, when $\tilde U$ is larger than $U_t$, the transition is
first order, and the critical temperature can be evaluated numerically.
Figure 2. shows the behavior of $T_c$ vs. $\tilde U$ in the two regions.
The value $\tilde U=4$ correspond to the vanishing of both the critical
temperature and ${c_0}$.

Figure 2. can be compared with the rough estimate of the critical temperature
of the long-range order phase whose existence is proved for the
Falicov-Kimball model in [8]. If one assumes that the phase with $c\neq 0$
could possibly be the long range order phase, the qualitative behavior of
$T_c$ is in agreement with that given by Kennedy and Lieb for large $U$,
whereas it is in contrast with the latter for vanishing $U$. One should
notice however that our approximation is expected to be more realistic
for finite $U$.

Finally, in figure 3. we give the behavior of $\displaystyle{{c}}$ vs.
$T$ for various fillings, still for a symmetric state ($n=d$). The figure
shows that the transition is present at different fillings, again in agreement
with the features of the long-range ordered phase described in [8].
\bigskip
{\bf 4. Conclusions}
\medskip
In this paper we have proposed an improvement of the Fermi-linearization
technique for electron systems, based on the requirement that the coefficients
of the linearized operators are Clifford-like variables, with their
anticommutator equal to an unknown constant $c$. As an example, we applied such
method to the Falicov-Kimball model, also giving the self-consistency equation
which determines the unknown. The latter turned out to behave as a true order
parameter, which at $T=0$ and at half-filling was shown to coincide
with the discontinuity in the single particle occupation number at the Fermi
level in the Gutzwiller approximation to the Hubbard model. The behavior of
$c$ was thus related to
the existence of a metal-insulator transition, which again was shown to
coincide at $T=0$ with that hypotized by Brinkman and Rice$^{[12]}$.

The above results suggest that our approximation could be a natural extension
of the Gutzwiller approach to the case $T\neq 0$. They also provide a physical
interpretation to the method, which consists in replacing the hopping term by a
term which locally still allows to create and annihilate electrons, but with an
amplitude proportional to the discontinuity in the single particle average
number at the Fermi surface.

Moreover, as opposite to the case in which the coefficients of the linearized
operators were Grassmann variables, the present approximation produces
non-trivial results even in the case $U=0$.

This paper was intended as a presentation of the method, and little
efforts were devoted to the numerical results in the various cases.
Nevertheless, in view of the promising results obtained, work is in progress
in order both to provide a
complete phase space at $T=0$ and to discuss the $T\neq 0$ behavior of the
physical quantities. We expect that also in this case the use
of a cluster Bethe version of our approximation$^{[3]}$ should give more
accured quantitative results.

Finally, let us stress that the method is of further generality. In particular,
we expect that it can be straightforwardly applied to the conventional Hubbard
model, as well as to generalized Hubbard models which have been proposed for
the study of high-$T_c$ superconductivity.
\bigskip
{\cpp The authors gratefully acknowledge stimulating discussions and a careful
reading of the manuscript by Mario Rasetti. One of the authors (A.M.) also
thanks Dan Mattis for letting her know about ref. [4].}

\vfill\eject
\line {{\bf References} \hfill}
\vskip .07 truein
\item{[1]} A. Montorsi, M. Rasetti, and A.I. Solomon, {\sl Phys. Rev. Lett.}
{\bf 59}, 2243 (1987)
\vskip .05 truein
\item{[2]}  A. Montorsi, M. Rasetti, and A.I. Solomon, {\sl Int. J. Mod.
Phys.} B{\bf 3}, 247 (1989)
\vskip .05 truein
\item{[3]} M. Bechi, R. Livi, and A. Montorsi, {\sl Int. J. Mod. Phys.}
B{\bf 5}, (1991)
\vskip .05 truein
\item{[4]} P. Prabasaj, and D.C. Mattis, {\sl Phys. Rev.} B{\bf 44}, 2384
(1991)
\vskip .05 truein
\item{[5]} L.M. Falicov, J.C. Kimball, {\sl Phys. Rev. Lett.} {\bf 22}, 997
(1969)
\vskip .05 truein
\item{[6]}J. Hubbard, {\sl Proc. R. Soc. London}, Sec. {\sl A} {\bf 276},
238 (1963) and {\bf 277}, 237 (1964); M.C. Gutzwiller {\sl Phys. Rev. Lett.}
{\bf 10}, 159 (1963)
\vskip .05 truein
\item{[7]} U. Brandt and C. Mielsch, {\sl Z. Phys. B -- Condensed Matter }
{\bf 75}, 365 (1989); {\bf 79}, 295 (1990)
\vskip .05 truein
\item{[8]} T. Kennedy, and E.H. Lieb, {\sl Physica} {\bf 138}A, 320 (1986)
\vskip .05 truein
\item{[9]} L. Gruber, J. Wanski, J. Jedrzejiski, and P. Lemberger,
{\sl Phys. Rev.} B{\bf 41}, 2198 (1990); P. Lemberger, {\sl J. Phys. A: Math.
Gen.} {\bf 25}, 715 (1992)
\vskip .05 truein
\item{[10]} V. Janis, {\sl Z. Phys. B -- Condensed Matter} {\bf 83}, 227 (1991)
\vskip .05 truein
\item{[11]} M. Gutzwiller, {\sl Phys. Rev. }{\bf 137}, A1726 (1965)
\vskip .05 truein
\item{[12]} W.F. Brinkman, and T.M. Rice, {\sl Phys. Rev.} B{\bf 2}, 4302
(1970)
\vfill\eject
{\bf Figure captions}
\medskip
\item{Fig. 1.} ${c}$ vs $\displaystyle{{kT}\over {qt}}$ at different $\tilde U$
values: $\tilde U = 0$ (continuous line), $\tilde U = 1$ (dashed line),
$\tilde U = 2$ (dotdashed line).
\smallskip
\item{Fig. 2.} $\displaystyle{{kT_c}\over {q t}}$ vs $\tilde U$: continuous
line represents second order transition, dashed line first order transition.
\smallskip
\item{Fig. 3.} ${c}$ vs $\displaystyle{{kT}\over {qt}}$ at $\tilde U =1$ and
different fillings, in the neutral case ($n=d$): $n+d=1$ (continuous line),
$n+d=.8$ (dashed line), $n+d=.6$ (dotdashed line).
\vfill\eject
\bye